\documentclass[sigconf]{acmart}
\usepackage{amsmath}
\usepackage{siunitx}
\usepackage{subcaption}
\usepackage{tcolorbox}
\tcbset{colback=white}
\graphicspath{{figures/}}
\copyrightyear{2020}
\acmYear{2020}
\acmConference[MSR '20]{17th International Conference on Mining Software
Repositories}{October 5--6, 2020}{Seoul, Republic of Korea}
\acmBooktitle{17th International Conference on Mining Software Repositories
(MSR '20), October 5--6, 2020, Seoul, Republic of Korea}
\renewcommand{\rq}[1]{\textbf{RQ{#1}}}
\newcommand{\rql}[1]{\textbf{Research Question {#1}:}}

\setcopyright{none}
\renewcommand\footnotetextcopyrightpermission[1]{} % removes footnote with conference information in first column

\begin{document}

\title[A Case Study of OpenSSL]{The Impact of a Major Security Event on an Open Source Project: The Case of OpenSSL}

\author{James Walden}
\email{waldenj@nku.edu}
\affiliation{%
    \institution{Northern Kentucky University}
    \streetaddress{1 Nunn Drive}
    \city{Highland Heights}
    \state{KY}
    \postcode{41076}
}

% Spaces between sections vanish with latexmk, but appear with pdflatex
\begin{abstract}
    {\it Context:} The Heartbleed vulnerability brought OpenSSL to international attention in 2014. The almost moribund project was a key security component in public web servers and over a billion mobile devices. This vulnerability led to new investments in OpenSSL.
    %Heartbleed led the Linux Foundation to create the Core Infrastructure Initiative (CII) to provide funding and advice to improve the security of key open source projects.
    % OpenSSL was one of the first projects funded.
    % The open source community immediately responded in two ways. The first was the creation of two forks of OpenSSL to improve code quality and security. The second was the 

    {\it Objective:} The goal of this study is to determine how the Heartbleed vulnerability changed the software evolution of OpenSSL. We study changes in vulnerabilities, code quality, project activity, and software engineering practices.

    {\it Method:} We use a mixed methods approach, collecting multiple types of quantitative data and qualitative data from web sites and an interview with a developer who worked on post-Heartbleed changes. We use regression discontinuity analysis to determine changes in levels and slopes of code and project activity metrics resulting from Heartbleed.

    {\it Results:} The OpenSSL project made tremendous improvements to code quality and security after Heartbleed. By the end of 2016, the number of commits per month had tripled, 91 vulnerabilities were found and fixed, code complexity decreased significantly, and OpenSSL obtained a CII best practices badge, certifying its use of good open source development practices.

% This case study demonstrates that the number of vulnerabilities and the rate at which they are discovered can change dramatically. This means that vulnerability count is {\bf not} a useful indicator of project security. Our results indicate that project activity and practices {\it may} be better predictors of project security. The CII badge {\tt may} be a good indicator of project security.
    % Our contribution is the first case study of the impact of a security event on open source software evolution. This study leads to a better
    % understanding of software evolution, and thus has the potential to
    % advance the practice of software development and maintenance.
    % It is one of a few studies that examine the impact of an external event on software evolution.
    {\it Conclusions:} The OpenSSL project provides a model of how an open source project can adapt and improve after a security event. The evolution of OpenSSL shows that the number of known vulnerabilities is not a useful indicator of project security. A small number of vulnerabilities may simply indicate that a project does not expend much effort to finding vulnerabilities. This study suggests that project activity and CII badge best practices may be better indicators of code quality and security than vulnerability counts.
%Instead of OpenSSL being replaced with another TLS implementation as multiple authors recommended, the project flourished after Heartbleed, becoming more active and improving in software practices, code quality, and security.

\end{abstract}




\maketitle
% Add for arxiv.org preprint
\pagestyle{plain} % removes running headers

\section{Introduction}

The OpenSSL project came to the attention of the world with the Heartbleed vulnerability on April 7, 2014. The open source cryptographic library was widely used to secure communications using the Transport Layer Security (TLS) protocol. An estimated 24-55\% of popular web sites using TLS were exposed to Heartbleed attacks~\cite{durumeric2014matter}, which allowed attackers to remotely access private keys and passwords. The number of client devices impacted was much larger, as OpenSSL was the default cryptographic library for Android devices, and more than one billion Android devices shipped in 2014~\cite{cnet}.

At the time of Heartbleed, the OpenSSL project was largely inactive, except for an ever-growing number of unaddressed issues. The project had no full time developers, and the OpenSSL Software Foundation received about \$2,000 per year in donations~\cite{money}. There were no policies for handling issues or vulnerabilities. OpenSSL had no release plan, and the project was still supporting version 0.9.8, which had been released in 2005. Project source code was complex and difficult to understand and maintain, while the project team was small with static membership~\cite{rsa16}.

The goal of this case study is to understand how the OpenSSL project responded to the Heartbleed vulnerability. We build this understanding with a mixed methods approach. We collect qualitative data from web sites and a developer interview, and we use quantitative data in the form of software metrics, including both project activity and code metrics. Our choice of metrics was guided in part by Lehman's laws of software evolution~\cite{lehman1996laws}. In particular, we analyze the first (continuing change) law using project activity metrics, the second law (increasing complexity), using code complexity metrics, and the sixth law (continuing growth) using code size metrics. We use regression discontinuity analysis to determine changes in levels and slopes of software metrics resulting from Heartbleed.

We believe this is the first study of the impact of a major security incident on software evolution. Our primary contribution is an understanding of how a project can recover from a major security incident. We learn that the number of reported vulnerabilities is a poor indicator of security. Project activity and software engineering practices required by the CII best practices badge~\footnote{\url{https://bestpractices.coreinfrastructure.org/}} may be better indicators of project security. Finally, we provide a replication package that includes both the data used in this paper and the code used to collect and analyze the data~\cite{msr2020replication}. Code to build the tables and figures in this paper is also included.

\begin{table*}

\caption{\label{tab:versions}OpenSSL Versions}
\centering
\begin{tabular}[t]{llllrrrrl}
\toprule
Version & Release Date & Last Release & Last Release Date & Release Count & Lifespan & End of Life & Code Size & Vulnerabilities\\
\midrule
0.9.1 & 1998-12-21 & 0.9.1c & 1998-12-23 & 1 & 2 days &  & 39,049 & \\
0.9.2 & 1999-03-22 & 0.9.2b & 1999-04-06 & 1 & 15 days &  & 40,112 & \\
0.9.3 & 1999-05-25 & 0.9.3a & 1999-05-27 & 2 & 2 days &  & 41,222 & \\
0.9.4 & 1999-08-09 & 0.9.4 & 1999-08-09 & 1 & 0 days &  & 41,973 & \\
0.9.5 & 2000-02-28 & 0.9.5a & 2000-04-01 & 2 & 33 days &  & 44,594 & \\
\addlinespace
0.9.6 & 2000-09-24 & 0.9.6m & 2004-03-17 & 14 & 1270 days &  & 47,025 & 10\\
0.9.7 & 2002-12-31 & 0.9.7m & 2007-02-23 & 14 & 1515 days &  & 60,442 & 9\\
0.9.8 & 2005-07-05 & 0.9.8zh & 2015-12-03 & 35 & 3803 days &  & 73,837 & 103\\
1.0.0 & 2010-03-29 & 1.0.0t & 2015-12-03 & 21 & 2075 days &  & 91,765 & 92\\
1.0.1 & 2012-03-14 & 1.0.1u & 2016-09-22 & 22 & 1653 days & 2016-12-31 & 100,415 & 127\\
\addlinespace
1.0.2 & 2015-01-22 & 1.0.2u & 2019-12-20 & 22 & 1793 days & 2019-12-31 & 107,862 & 94\\
1.1.0 & 2016-08-25 & 1.1.0l & 2019-09-10 & 13 & 1111 days & 2019-09-11 & 100,957 & 30\\
1.1.1 & 2018-09-11 & 1.1.1d & 2019-09-10 & 5 & 364 days & 2023-09-11 & 116,486 & 8\\
\bottomrule
\end{tabular}
\end{table*}

\section{Context}

The first version of OpenSSL was released on December 23, 1998. It was based on a fork of the SSLeay project. The project name comes from the Secure Sockets Layer (SSL) protocol, which has been deprecated in favor of Transport Layer Security. OpenSSL consists of two libraries, which provide support for TLS and general cryptographic algorithms respectively, and a command line tool, {\tt openssl}, which can be used for encryption, decryption, and certificate generation.

Table~\ref{tab:versions} summarizes the major versions of OpenSSL, including lifespan, minor releases, code size (number of statements), and number of vulnerabilities reported per release. The version number format is three digits followed by an optional letter, indicating minor releases. Prior to Heartbleed, OpenSSL had no release policy. Therefore, earlier versions have no end of life date. Lifespan is the period between the first and last release of a numerical version. 

Vulnerability reporting started in 2002, so earlier versions have no vulnerabilities associated with them. Vulnerabilities often affect multiple versions of OpenSSL, so the total number of vulnerabilities is smaller than the sum of the vulnerabilities in the table.

Heartbleed focused intense scrutiny on OpenSSL. Two approaches to OpenSSL were advocated: replacement or repair. Two forks were created shortly after Heartbleed as potential replacements. Google's fork, BoringSSL~\cite{boringssl}, focused on supporting Android and Chrome. The OpenBSD project's fork was called LibreSSL~\cite{libressl30days} and focused on improving security and code quality.

The repair approach was supported by the Core Infrastructure Initiative (CII), which was started by the Linux Foundation as a response to Heartbleed. The purpose of the CII was to fund and support open-source projects that are critical to the functioning of the Internet~\cite{ciipress}. OpenSSL was among the first projects funded. CII funded two full-time developers and a code audit. An additional two developers were funded by donations~\cite{rsa16}. CII supported the project's first face-to-face meeting in late 2014, during which project members drafted major policies, including a release strategy, coding style guide, and security policy.

\section{Related Work}
Lehman's laws~\cite{lehman1996laws} provide a framework for understanding software evolution that has been widely studied. Previous studies have operationalized these laws terms of software metrics in multiple ways~\cite{neamtiu2013towards,gonzalez2014studying,li2017evolution}. We examine the first, second, and sixth laws in this work. Multiple studies using a variety of metrics have found that these three laws hold for most but not all open source projects studied~\cite{godfrey2001growth,paulson2004empirical,wu2004linker,izurieta2006evolution,neamtiu2013towards}. 

To study Lehman's second law, we use code complexity metrics. McCabe developed his cyclomatic complexity metric as a quantitative measure of which software modules are difficult to maintain or test~\cite{mccabe}. Halstead developed complexity metrics for similar purposes~\cite{halstead1977elements}. Midha et al.~\cite{midha2010improving} found that the number of bugs in software increased with increasing cyclomatic complexity, while Gill and Kemerer~\cite{gill1991cyclomatic} found that complexity density (the ratio of cyclomatic complexity to lines of code) was a useful predictor of maintenance productivity. Halstead's and McCabe's complexity metrics have been used as predictors in defect prediction~\cite{menzies2002metrics, menzies2006data} and vulnerability prediction~\cite{shin2010evaluating} models.

We use code size metrics to study Lehman's sixth law. Code size metrics have been used in multiple studies of Lehman's laws~\cite{gonzalez2014studying,li2017evolution}. Jimenez et al. describe characteristics of vulnerable files in OpenSSL using code size and code complexity metrics~\cite{jimenez2016empirical}, but do not use them to study software evolution. We compute both code size and code complexity metrics using the {\tt cqmetrics} tool that Spinellis used to study the evolution of the Unix operating system~\cite{spinellis2016evolution}. 
% Mockus, Fielding, and Herbsleb studied open source development using case studies of Apache and Mozilla~\cite{mockus2002two}, but did not investigate the security of those projects. There are no similar case studies of open source development for OpenSSL. 

In addition to code size and complexity, we examine code style and language feature use. Following a consistent coding style may be an important aspect of readability and maintainability of software~\cite{oraclejava,smit2011maintainability}. Programming languages like Java~\cite{oraclejava} and Python~\cite{pep8} have style guides, while organizations like Google~\cite{googlestyle} publish style guides for a variety of languages. Smit et al.~\cite{smit2011maintainability} found that in the absence of automated style checkers, the number of style violations grows in a linear relationship with code size.

While the idea that use of {\tt goto} is harmful~\cite{dijkstra1968go} has been widely heard by programmers, an empirical study of GitHub projects suggested that use of {\tt goto} for specific purposes like error handling was considered good practice by open source developers~\cite{nagappan2015empirical}. On the other hand, use of C preprocessor conditionals for portability across different architectures has continued to receive blame for making code difficult to read and maintain~\cite{ifdef,miller2005secure}.

While there are many studies of open source evolution, few studies examine the impact of an external change on software evolution. We use a regression discontinuity design~\cite{reichardt2019quasi,jacob2012practical} (RDD) to assess the impact of Heartbleed on the evolution of OpenSSL. This methodology has recently begun to be used in empirical software engineering analyses of time series data. Zhao et al. used RDD to evaluate the impact of adopting continuous integration on other software development practices~\cite{zhao2017impact}, while Trockman et al. used RDD to test whether repository badges were reliable signals of software quality~\cite{trockman2018adding}. Zimmermann and Art{\'\i}s evaluated the impact of switching bug trackers on a single project with a RDD model~\cite{zimmermann2019impact}.

Durumeric et al. measured the reaction to Heartbleed, finding that Alexa Top 100 sites patched within 48 hours, while less popular sites took longer to deploy patches~\cite{durumeric2014matter}. Kupsch and Miller discuss the difficulty software security tools have in finding vulnerabilities like Heartbleed~\cite{kupsch2014software}. Wheeler describes software engineering practices and technologies that could find vulnerabilities like Heartbleed, including simplifying the code, fuzzing with address checking, and thorough security-focused testing~\cite{wheeler2014prevent}. We will examine how OpenSSL adopted some of those approaches below. 

\section{Research Questions}

% git log 96db9023b881d7cd9f379b0c154650d6c108e9a3 -1 --stat
% While the only response to many vulnerabilities is a single code commit and a new minor release, the response to Heartbleed was more substantial.
% The CII funded additional developers and a code audit of OpenSSL, along with creating the CII badge project ~\footnote{\url{https://bestpractices.coreinfrastructure.org/}} to certify open source projects that followed good open source development practices. This investment helped OpenSSL transition from an almost moribund project to an active project with a focus on improving code quality and security. Studying the impacts of this transition can help us understand the impact of investing in open source projects.

We begin our study by examining the security of OpenSSL. We examine the number and severity of vulnerability reports over time to understand how the Heartbleed vulnerability affected vulnerability reporting. Our first research question is:

% Parameters in [] seem to affect all following tcolorboxes too
\begin{tcolorbox}[boxsep=0mm]
\rql 1 How did the number and severity of reported vulnerabilities change after Heartbleed?
\end{tcolorbox}

We expect to see an increased number of vulnerability reports after Heartbleed, as more effort was devoted to finding security issues. Such an increase is an indicator of improving project security as vulnerabilities are found and remediated. 

% ./Configure LIST | wc -l gives 150 platforms
OpenSSL has been under development for over two decades, a substantial span of time in which open source development has greatly changed. Lehman's sixth law of software evolution states that software continually grows in functionality to maintain user satisfaction~\cite{lehman1996laws}. Adding new functionality typically requires adding more code to a project. Therefore, we assess the size of OpenSSL to see if this law holds after a major security incident. It is worth noting that both of the post-Heartbleed forks of OpenSSL began development by removing older cryptographic algorithms and support for a variety of computing environments~\cite{boringssl, libressl30days}. Our second research question is:

\begin{tcolorbox}[boxsep=0mm]
\rql 2 How did OpenSSL change in size after Heartbleed?
\end{tcolorbox}

% FIXME: Add complexity effects on open source migrants, reducing ndevelopers?
The quality of OpenSSL code before Heartbleed was perceived as poor~\cite{libressl30days}. It was difficult to read and maintain, discouraging new contributors from working on the project~\cite{rsa16}. Therefore, we want to measure code quality using metrics that impact the readability and maintainability of code, such as code complexity metrics.

Lehman's second law of software evolution states that program complexity increases over time unless work is done to prevent that~\cite{lehman1996laws}. With two decades of history, there has been ample time for OpenSSL's code to increase in complexity.
While some complexity is necessary for cryptographic code, unnecessary complexity can accumulate over time. The complexity of OpenSSL code was implicated as one of the causes of the Debian project accidentally breaking the OpenSSL pseudo-random number generator in 2006~\cite{cox}.

\begin{tcolorbox}[boxsep=0mm]
\rql 3 How did the complexity of OpenSSL source code change after Heartbleed?
\end{tcolorbox}

% See spinellis2016evolution section 4: related work
% See references in https://www.simula.no/sites/default/files/publications/Simula.SE.683.pdf
% An important part of code maintenance is reading and understanding code. 
As code style and the use of certain programming language features can affect the readability and maintainability of software, we also study these characteristics of the OpenSSL code base. We examine the consistency of stylistic choices like bracket placement and indentation, and we study the use of certain language features in our study of coding style, such as the C preprocessor and the {\tt goto} statement. 

\begin{tcolorbox}[boxsep=0mm]
\rql 4 How did the coding style of OpenSSL source code change after Heartbleed?
\end{tcolorbox}

Lehman's first law focuses on continuing change of a software project. One of the major problems with OpenSSL at the time of Heartbleed was insufficient developer activity to address technical debt. The project team was also small and included no full time developers, while team membership was static~\cite{rsa16}. The number of contributions from outside developers was small. There were no guidelines for contributing to the project, while the source code was difficult to understand and maintain, discouraging new contributors. 

The time for the developers to respond to issues was high, and most issues were not addressed. The number of open, unaddressed issues had grown steadily, reaching almost 1500 at the time of Heartbleed~\cite{openssl_issues}. 

To measure the change in project activity, we count the number of commits before and after Heartbleed and measure the number of commits per month. We also want to determine whether the OpenSSL project was able to grow its development team and accept a substantial number of outside contributions in the wake of Heartbleed, so we measure the number of authors.

\begin{tcolorbox}[boxsep=0mm]
\rql 5 How did the number of commits and the number of authors change after Heartbleed?
\end{tcolorbox}

We also study the software engineering practices that led to changes in project activity, code quality, and security. As the only requirements and design documents available for OpenSSL are for the forthcoming 3.0.0 release, we focus on implementation and testing activities that have been identified as best practices by the CII Best Practices Badge project~\cite{ciibadge_criteria}. 

% Open source development has changed over the two decades of OpenSSL's history. When OpenSSL began, projects used centralized version control systems and open source contributors typically submitted patches to core developers via e-mail for inclusion. In the current era, projects typically use distributed version control systems and receive contributions via pull requests. At the time of Heartbleed, OpenSSL had recently transitioned to the {\tt git} distributed version control system but still followed the older pattern of submitting code patches via e-mail.

\begin{tcolorbox}[boxsep=0mm]
\rql 6 How did software engineering practices change after Heartbleed?
\end{tcolorbox}

Finally, we want to determine if the changes in OpenSSL were sustained well after the discovery of the Heartbleed vulnerability. To address this question, we will not only examine trends in our code and activity metrics through the end of 2019 but we will also examine external assessments of the OpenSSL project.

\begin{tcolorbox}[boxsep=0mm]
\rql 7 Are the changes in OpenSSL's security, source code, and software engineering practices sustained five years after Heartbleed? 
\end{tcolorbox}

\section{Data}

We collected data on the OpenSSL project from a variety of sources, including the project web site, vulnerabilities list, and GitHub repository. We also interviewed one OpenSSL developer.

\textit{Vulnerabilities:} Vulnerability data was collected from the OpenSSL vulnerabilities list~\footnote{\url{https://www.openssl.org/news/vulnerabilities.html}}. We count vulnerabilities using unique Common Vulnerabilities and Exposures (CVE) identifiers.

\textit{Code Metrics:} The source code of OpenSSL was obtained from the project's GitHub repository~\footnote{\url{https://github.com/openssl/openssl}}. We used the {\tt cqmetrics} package~\footnote{\url{https://github.com/dspinellis/cqmetrics}} to compute code metrics, including size, complexity, language feature use, and style metrics. This open source tool was chosen in part because of its prior use in computing metrics on many versions of Unix released over several decades~\cite{spinellis2016evolution}. For our monthly time series data, we compute code metrics on the first commit made during a month.

\textit{Project Activity:} Metrics on project activity, such as the number of authors and commits were obtained from the project's GitHub repository using PyDriller~\cite{Spadini2018}. We compute authors per month as the number of unique author names in commits made during a month.

\textit{Software Engineering Practices:} We reviewed the OpenSSL web site, mailing lists, and GitHub repository for information on changes in software engineering practices. We collected data on unit testing from the project GitHub repository and from test coverage data reported via {\tt coveralls.io}.

\textit{Interviews:} Our interview requests received a single response from an OpenSSL team member who worked on the project immediately after Heartbleed. We interviewed Rich Salz of Akamai, who joined the project a couple of months after Heartbleed. He helped change software engineering processes, including moving to a GitHub pull request work flow, transitioning issue tracking to GitHub, and adopting continuous integration.
%Our questions focused on changes in software engineering practices and the reasons for those changes.

\section{Methods}

We use data visualization and statistical modeling to discover changes in code metrics and project activity. In particular, we use a regression discontinuity design~\cite{reichardt2019quasi,jacob2012practical} approach to analyze time series of code and project activity metrics. RDD allows us to determine whether changes in metrics occurred and to measure the effect size of those changes.

Regression discontinuity design is a rigorous quasi-experimental approach for analyzing the casual effect of an intervention. It is based on the idea that in the absence of the intervention, the observed trend before the intervention would continue afterwards.
In such a design, observations are assigned to an intervention condition based on a cutoff score. In time series analysis, the cutoff is a date. The cutoff date can be determined by visually inspecting the time series for discontinuities or by the date of a specific event. 
The effect of the intervention is estimated as a discontinuity between the intervention groups before and after the cutoff date. When using a linear regression model, the intervention effect can include both a change in level and a change in slope at the cutoff. 
%In this study, we will use different cutoffs for different research questions, due to different events affecting different parts of the project.

Regression discontinuity can be performed with a global or local regression approach. In global regression approaches, the entire data set is used to fit the model. In local regression models, the model is fitted using a subset of the time series, with some data on each side on the cutoff. This amount is called the bandwidth. RDD works better with an equal amount of data before and after the cutoff. While the global approach offers greater precision, it includes data far from the cutoff, which may be influenced by trends other than the intervention being studied. Given the extensive history of OpenSSL before Heartbleed, we choose a local regression approach.

We use the following regression discontinuity model equation to estimate changes in level and trend in code metrics after the beginning of major post-Heartbleed work:
\begin{equation*}
    y_i = \alpha + \tau D + \beta_1 (t_i - c) + \beta_2 D (t_i - c) + \epsilon_i
\end{equation*}
where D is a function that represents the discontinuity 
\begin{equation*}
D = \begin{cases} 
      0 & t_i < c \\
      1 & t_i \geq c \\
   \end{cases}
\end{equation*}
In the equation above, $c$ is the cutoff date, when code metrics showed substantial changes. The coefficient $\alpha$ represents the level of the regression line before the cutoff, while $\tau$ represents the change in level after the cutoff. The sum $\alpha + \tau$ is the level of the regression line after the discontinuity.

The response variable $y_i$ is the value of a particular code metric at time $t_i$. The variable $t_i$ represents time in months. The coefficient $\beta_1$ is the slope of the regression line before the cutoff, while $\beta_2$ is the change in slope after the cutoff. The sum $\beta_1 + \beta_2$ is the slope of the regression line after the cutoff. The variable $\epsilon_i$ represents the error at time $t_i$.

In our analysis of code metrics, we build RDD models with a cutoff of February 2015. This cutoff was initially identified visually. It is clear to see in the plots of nesting depth and style inconsistency in Figure~2, which approach a step function that transitions from one interval to the other in this month. While Heartbleed was the initial impetus for changing OpenSSL, large changes to the code base began in February 2015. Investigation of the project web site and mailing lists reveal that the time between Heartbleed and our cutoff date was spent building the team, developing policies, and planning how to change the code base.

Confirming our choice of cutoff, the OpenSSL project blog published an article about reformatting the entire code base to meet the project's new coding style guidelines in February 2015~\cite{openssl_reformat}. Additional code cleanup was performed in the following months, described in another blog entry published in July 2015~\cite{openssl_cleanup}. We choose a bandwidth of 25 months on each side of the cutoff date, so that we have a sufficient number of data points for our model without including data points that are so far from the cutoff that they are influenced by factors other than Heartbleed.

\section{Results}

\subsection{Vulnerabilities}

\begin{figure}[ht!]
    \centering
    \includegraphics[width=\linewidth]{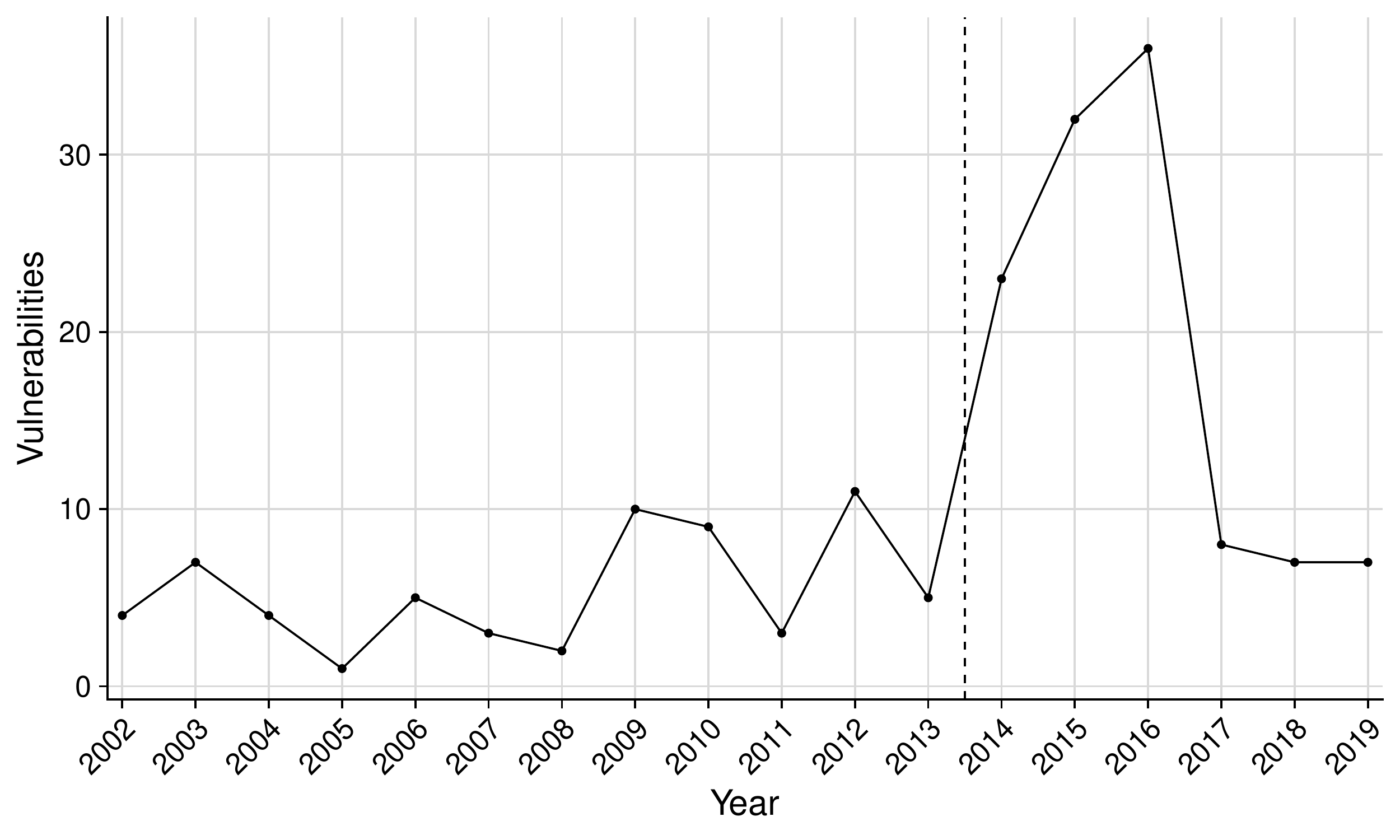}
    \caption{OpenSSL Vulnerabilities Reported by Year}
    \label{fig:vulns}
    \Description{Time series graph of OpenSSL vulnerabilities.}
\end{figure}

% The project quickly published a security policy on vulnerability handling in September 2014.
% https://www.openssl.org/blog/blog/2015/09/01/openssl-security-a-year-in-review/
Our first research question focuses on security vulnerabilities. There have been 177 vulnerabilities reported in OpenSSL through the end of 2019. Figure~\ref{fig:vulns} shows the trend of the annual number of vulnerabilities. The date of Heartbleed is indicated by a dashed line. While 66 (37.3\%) vulnerabilities were reported in the approximately sixteen years before Heartbleed, 110 (62.1\%) vulnerabilities were reported in the five years after Heartbleed. 

% OpenSSL is not the only open source software in which vulnerability reports suddenly and dramatically increase. For another example, 133 vulnerabilities were published to the NVD for {\tt tcpdump} in a single year (2017), a number that dwarfs the total of 6 vulnerabilities that reported for this project between its inception in 1988 and that year.

We can see three eras of vulnerability reporting for the OpenSSL project in Figure~\ref{fig:vulns}: the pre-Heartbleed era, the high vulnerability reporting era from 2014 to 2016, and the modern era from 2017 to the present. Vulnerability statistics for the three eras are detailed in Table~\ref{tab:vulns}. 

The pre-Heartbleed era has the lowest number of vulnerabilities per year but with slightly higher severity scores (CVSSv2) than the other two eras. There is a dramatic increase in the number of vulnerabilities reported per year, from 4 to 30.33, in the high vulnerability era, with severities remaining on par with the pre-Heartbleed era. A majority (51.4\%) of all vulnerabilities were reported in the three year period between 2014 and 2016. Only five of the 91 vulnerabilities found in the 2014-2016 time period were found in code written after Heartbleed, so this era produced a substantial improvement in the security of OpenSSL.

In the modern era starting in 2017, the number of vulnerabilities reported per-year is almost twice as high as the annual count before Heartbleed, but the mean severity of vulnerabilities has declined, as has variability around that mean. However, the pre-Heartbleed mean number of vulnerabilities per year is computed across many more years than the mean for the modern era. If we restrict our view to the four years before Heartbleed, the mean number of vulnerabilities reported per-year is 7.0, which is quite close the modern era mean of 7.33. The mean CVSSv2 score for the four pre-Heartbleed years is 5.45 and the standard deviation is 0.84, so we still observe a decline in vulnerability severity between the late pre-Heartbleed era and the modern era.

% We cannot perform a reliable regression discontinuity analysis on vulnerability data, as we have only 18 data points. We also cannot construct a usable monthly trend, as the number of vulnerabilities reported in almost all months is zero.

\begin{table}

\caption{\label{tab:vulns}OpenSSL Vulnerabilities}
\centering
\begin{tabular}[t]{lrrr}
\toprule
 & Pre-Heartbleed & 2014-2016 & 2017-2019\\
\midrule
Vulnerability Count & 64 & 91 & 22\\
Vulnerability Percent & 36.2\% & 51.4\% & 12.4\%\\
Vulnerabilities per Year & 4.00 & 30.33 & 7.33\\
Average CVSS2 & 5.63 & 5.25 & 4.18\\
StdDev CVSS2 & 1.11 & 1.11 & 0.39\\
\bottomrule
\end{tabular}
\end{table}

\subsection{Code Size}

\begin{figure*}[htp]
    \begin{tabular}{cc}
        \begin{subfigure}{.5\textwidth}
            \centering
            \includegraphics[width=1.0\linewidth]{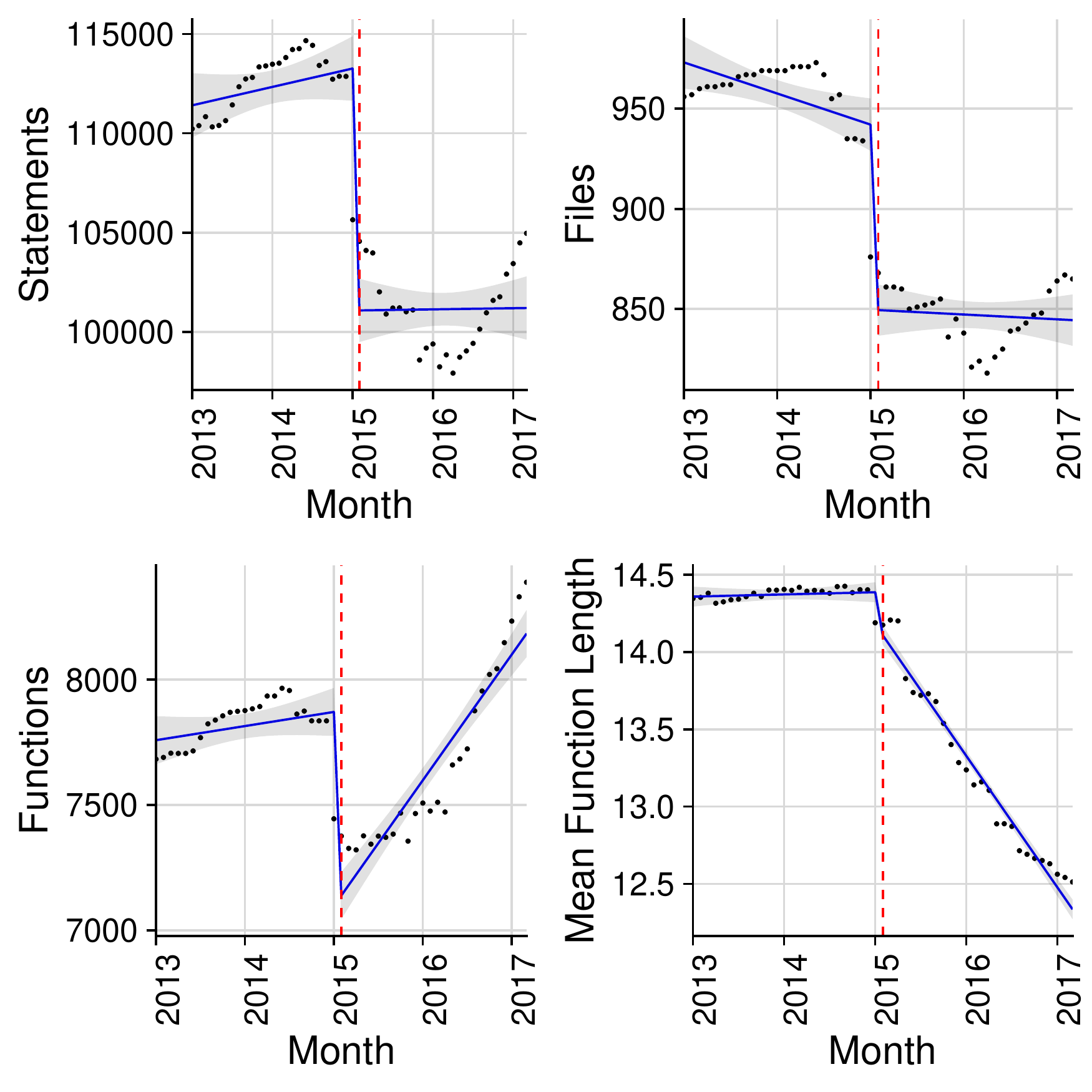}
            \caption{Code Size Metrics}
            \label{fig:code_size}
        \end{subfigure} &
        \begin{subfigure}{.5\textwidth}
            \centering
            \includegraphics[width=1.0\linewidth]{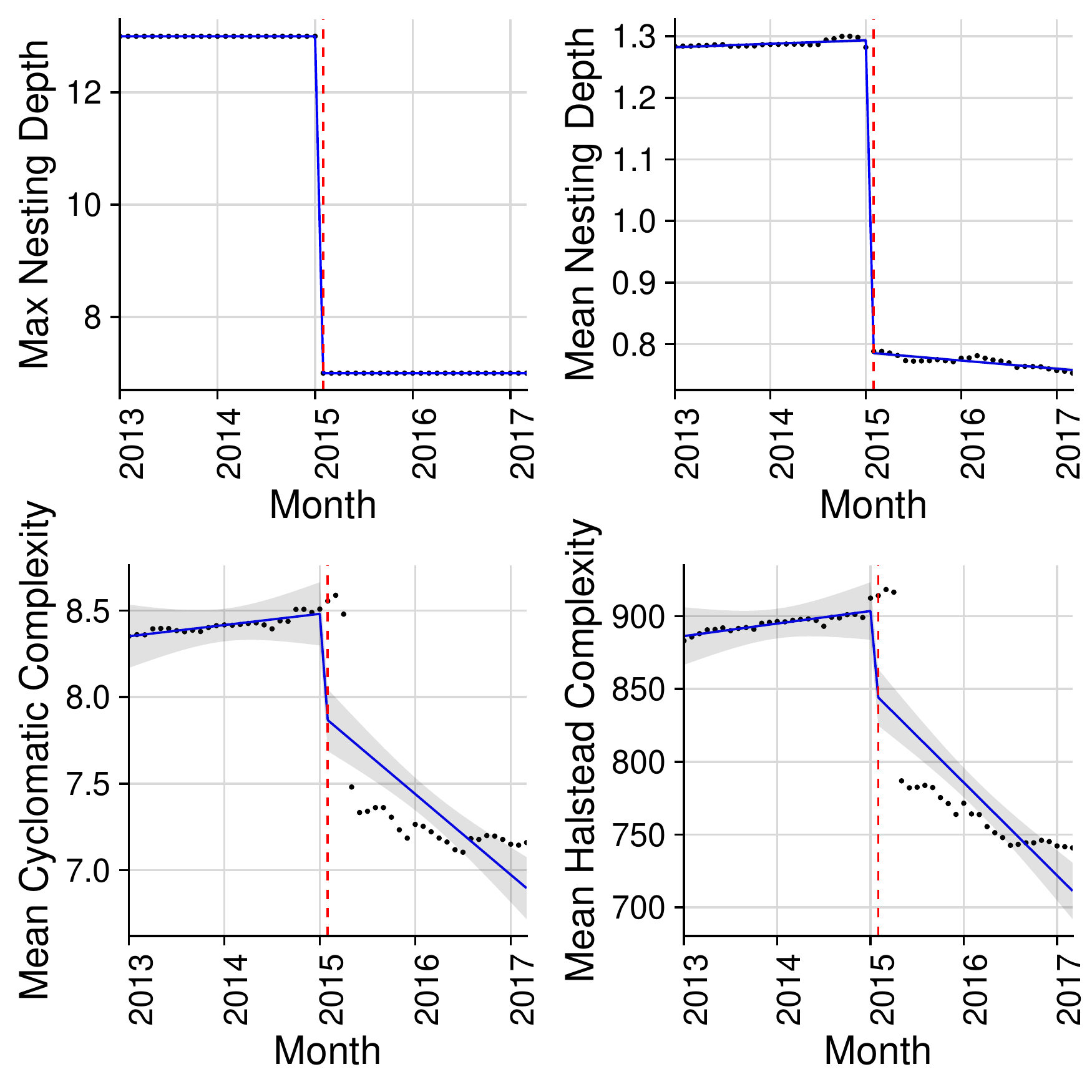}
            \caption{Code Complexity Metrics}
            \label{fig:code_complexity}
        \end{subfigure}\\

        \begin{subfigure}{.5\textwidth}
            \centering
            \includegraphics[width=1.0\linewidth]{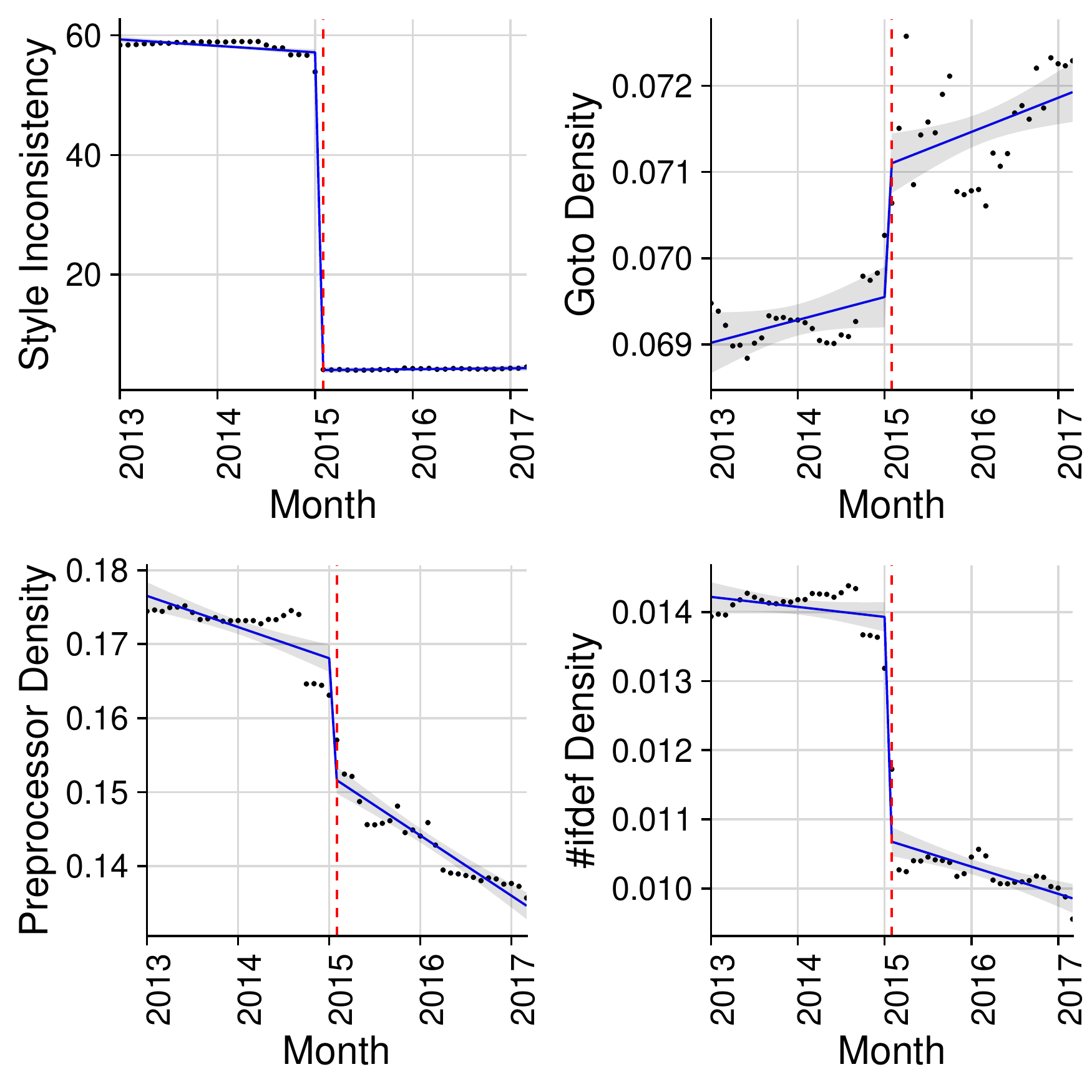}
            \caption{Code Style Metrics}
            \label{fig:code_style}
        \end{subfigure} &
        \begin{subfigure}{.5\textwidth}
            \centering
            \includegraphics[width=1.0\linewidth]{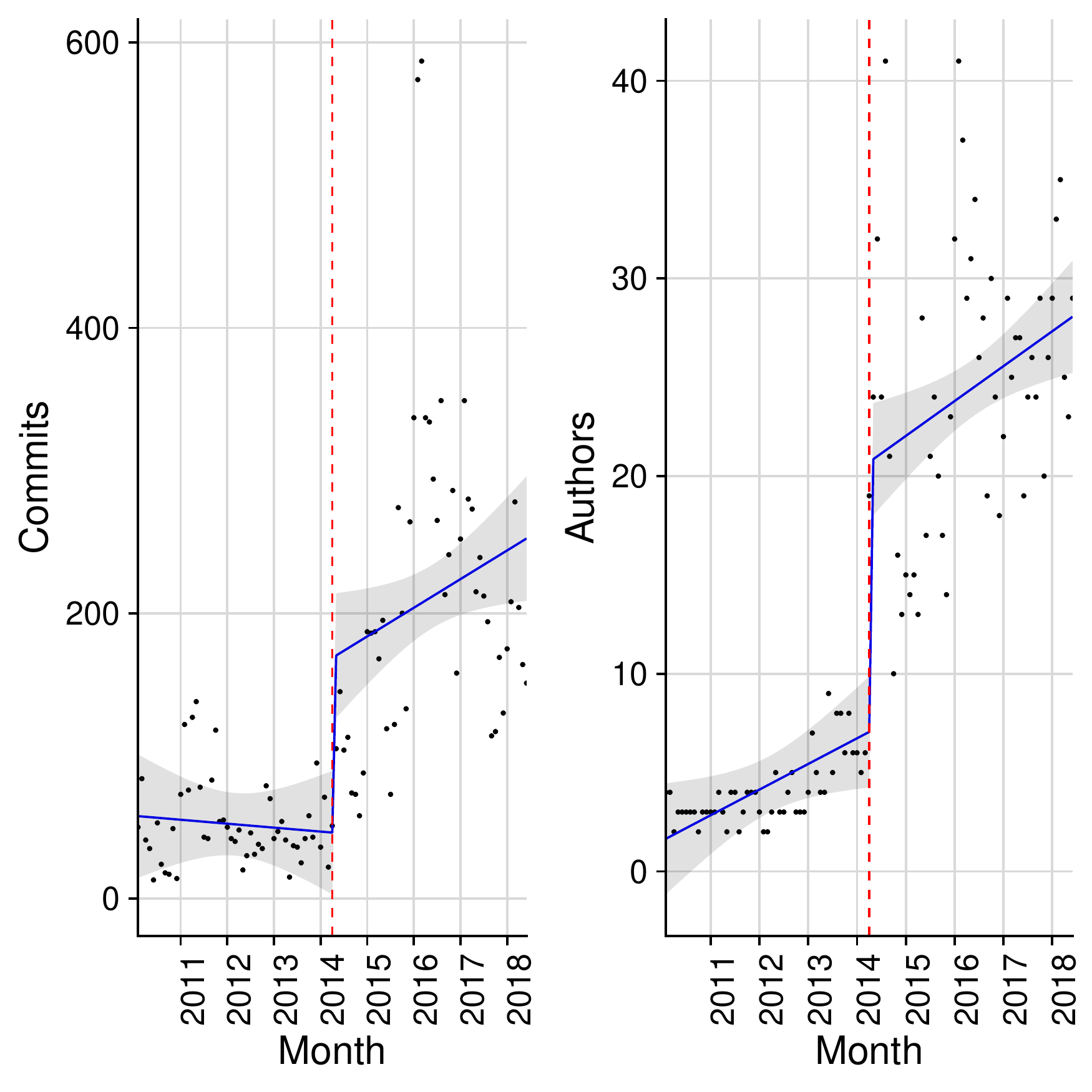}
            \caption{Project Activity Metrics}
            \label{fig:project_activity}
        \end{subfigure}\\
    \end{tabular}
    \label{rdd}\caption{Regression Discontinuity Models}
    \Description{Plots of RDD models of code metrics and project activity}
\end{figure*}

In this and the following three sections, we analyze monthly time series of the software metrics shown in Figure~2. We compute monthly code metrics by checking out the version of the code available from the first commit made during the month and running {\tt cqmetrics} on that version. Project activity metrics are computed using all commits during the month.

To study \rq 2, we analyze multiple code size metrics. Code size, both in terms of the number of files and number of statements, starts dropping shortly after Heartbleed, as can be seen in Figure~\ref{fig:code_size}. However, much larger changes begin during the code cleanup in February 2015, which we chose as the cutoff date for our RDD models. Table~\ref{tab:code_size} shows a significant decrease of 12,260 statements (10.8\% of pre-cutoff size) after code cleanup began in February 2015. We see a similar decrease in the number of files. Both models were good fits as measured using the adjusted $R^2$ metric, which is given in parentheses after the model name in Table~\ref{tab:code_size}.

% git log --diff-filter=D --since=2014-04-01 --before=2016-05-01 --oneline | wc -l
% git log --diff-filter=D --since=2014-04-01 --before=2016-05-01 --oneline --summary | egrep '\.[ch]$' | wc -l
To better understand these size changes, we examined 126 commits made in the 25 months after Heartbleed, in which the only file changes were deletions. A total of 315 C source and header files were deleted in those commits. We found multiple mentions of obsolete or old files, including support for obsolete platforms, old insecure protocols like SSLv2, and unmaintained demo code in developer commit messages of these commits. These messages point to the removal of files being an improvement of code quality, especially since support for old protocols exposed OpenSSL to protocol downgrade attacks like DROWN~\cite{aviram2016drown}. 

In Figure~\ref{fig:code_size}, we can see code size as measured in terms of the number of files and statements decreasing from Heartbleed until about the release of version 1.1.0 in August 2016. Version 1.1.0 was the only major release of OpenSSL to have a smaller size at the time of release than its predecessors. Code sizes measured by number of statements for each version are provided in Table~\ref{tab:versions}.
%However, code size reduction was a temporary reaction to Heartbleed. Version 1.1.1, released two years after 1.1.0, is 15\% larger than its predecessor, returning to adherence with Lehman's sixth law with the addition of new features, such as support for SHA3 and TLS 1.3.

\begin{table*}

\caption{\label{tab:code_size}Code Size Metrics Regression Discontinuity Models}
\centering
\begin{tabular}[t]{lrrrrrrrr}
\toprule
\multicolumn{1}{c}{} & \multicolumn{2}{c}{Statements (0.88)} & \multicolumn{2}{c}{Files (0.92)} & \multicolumn{2}{c}{Functions (0.79)} & \multicolumn{2}{c}{Mean Function Length (0.99)} \\
\cmidrule(l{3pt}r{3pt}){2-3} \cmidrule(l{3pt}r{3pt}){4-5} \cmidrule(l{3pt}r{3pt}){6-7} \cmidrule(l{3pt}r{3pt}){8-9}
  & Estimate & Std Error & Estimate & Std Error & Estimate & Std Error & Estimate & Std Error\\
\midrule
Pre-Intervention Level & 113340.58\rlap{\textsuperscript{***}} & 856.579 & 940.79\rlap{\textsuperscript{***}} & 6.893 & 7876.18\rlap{\textsuperscript{***}} & 50.579 & 14.389\rlap{\textsuperscript{***}} & 0.034\\
Change in Level & -12259.623\rlap{\textsuperscript{***}} & 1166.553 & -91.328\rlap{\textsuperscript{***}} & 9.387 & -737.394\rlap{\textsuperscript{***}} & 68.882 & -0.28\rlap{\textsuperscript{***}} & 0.046\\
Pre-Intervention Slope & 77.288 & 57.620 & -1.293\rlap{\textsuperscript{**}} & 0.464 & 4.697 & 3.402 & 0.001 & 0.002\\
Change in Slope & -72.429 & 79.191 & 1.093 & 0.637 & 37.12\rlap{\textsuperscript{***}} & 4.676 & -0.072\rlap{\textsuperscript{***}} & 0.003\\
\bottomrule
\multicolumn{9}{l}{RDD models for code size metrics, showing changes in each metric's mean and slope at February 2015. Each model's $R^2$}\\
\multicolumn{9}{l}{metric is shown in parentheses after metric name. Statistical significance is indicated by *** p < 0.001, ** p < 0.01, * p < 0.05}\\
\end{tabular}
\end{table*}

After February 2015, the number of functions had decreased significantly by 737 (9.3\%), but the number of functions being added per month was almost eight times the pre-cutoff number. To understand how the number of functions was increasing rapidly while the numbers of statements and files were declining, we examine mean function length. Function length transitions from an essentially flat pre-Heartbleed trend to a decreasing trend (-0.072 lines/month) after Heartbleed. This evidence suggests that the OpenSSL project began to factor their code into smaller functions after Heartbleed. Many programmers find that smaller functions are easier to understand~\cite{martin2009clean}. 
%Figure~\ref{fig:code_size} shows that after the drop, the number of functions quickly began increasing at a much higher rate than before Heartbleed (37 per month compared to 5 per month.) 

\subsection{Code Complexity}

Complex code is difficult to understand and maintain, so unnecessary complexity should be avoided. \rq 3 focuses on changes to OpenSSL's code complexity. We examine three types of code complexity metrics: nesting complexity, cyclomatic complexity, and Halstead complexity. Nesting complexity is the number of levels of nesting in source code. 

% \begin{figure}[ht]
%     \centering
%     \includegraphics[width=\linewidth]{code-complexity.pdf}
%     \caption{Code Complexity Models}
%     \label{fig:code_complexity}
%     \Description{Plots of RDD models of code complexity metrics}
% \end{figure}

\begin{table*}

\caption{\label{tab:code_complexity}Code Complexity Metrics Regression Discontinuity Models}
\centering
\begin{tabular}[t]{lrrrrrrrr}
\toprule
\multicolumn{1}{c}{} & \multicolumn{2}{c}{Max Nesting (1.00)} & \multicolumn{2}{c}{Mean Nesting (1.00)} & \multicolumn{2}{c}{Cyclomatic (0.85)} & \multicolumn{2}{c}{Halstead (0.87)} \\
\cmidrule(l{3pt}r{3pt}){2-3} \cmidrule(l{3pt}r{3pt}){4-5} \cmidrule(l{3pt}r{3pt}){6-7} \cmidrule(l{3pt}r{3pt}){8-9}
  & Estimate & Std Error & Estimate & Std Error & Estimate & Std Error & Estimate & Std Error\\
\midrule
Pre-Intervention Level & 13\rlap{\textsuperscript{***}} & 0 & 1.294\rlap{\textsuperscript{***}} & 0.002 & 8.486\rlap{\textsuperscript{***}} & 0.096 & 904.286\rlap{\textsuperscript{***}} & 10.425\\
Change in Level & -6\rlap{\textsuperscript{***}} & 0 & -0.508\rlap{\textsuperscript{***}} & 0.002 & -0.62\rlap{\textsuperscript{***}} & 0.131 & -60.071\rlap{\textsuperscript{***}} & 14.197\\
Pre-Intervention Slope & 0\rlap{\textsuperscript{*}} & 0 & 0\rlap{\textsuperscript{***}} & 0.000 & 0.005 & 0.006 & 0.717 & 0.701\\
Change in Slope & 0 & 0 & -0.002\rlap{\textsuperscript{***}} & 0.000 & -0.044\rlap{\textsuperscript{***}} & 0.009 & -6.032\rlap{\textsuperscript{***}} & 0.964\\
\bottomrule
\multicolumn{9}{l}{RDD models for complexity metrics, showing changes in each metric's mean and slope at February 2015. Each model's $R^2$}\\
\multicolumn{9}{l}{metric is shown in parentheses after metric name. Statistical significance is indicated by *** p < 0.001, ** p < 0.01, * p < 0.05}\\
\end{tabular}
\end{table*}

We study mean and maximum values of nesting complexity and mean values of cyclomatic and Halstead complexity. Mean complexity metrics are computed as a mean of the per-file mean values of these metrics for each month in the dataset. We construct four RDD models, which are summarized in Table~\ref{tab:code_complexity}. Figure~\ref{fig:code_complexity} shows both data and model fit for these four metrics.

The maximum depth of nesting in OpenSSL source files dropped suddenly from 13 to 7 in February 2015. This change affected 94 files, when the OpenSSL project reformatted its code base~\cite{openssl_reformat}. The RDD model of mean nesting depth shows a drop of approximately half a level from 1.29 to 0.79 in February 2015.

Mean cyclomatic complexity dropped by 7.3\% after February 2015. The trend of cyclomatic complexity evolution changes from a small increasing slope before intervention to a larger decreasing slope after the cutoff. The evolution of Halstead complexity follows the same pattern, with a drop of 6.6\% and a similar change in slope. These changes suggest that the post-Heartbleed code cleanup was effective at reducing the complexity of OpenSSL code and thus improving readability and maintainability.

\subsection{Coding Style}

\rq 4 focuses on coding style. Poor coding style can make code difficult to understand and maintain, increasing the likelihood of programmer errors and decreasing the likelihood of attracting contributors to the project.
Prior to the Heartbleed vulnerability, the OpenSSL project had no coding style guide, and different sections of code used a range of different styles~\cite{openssl_reformat}.
In January 2015, the project published a coding style guide~\cite{openssl_style}. In early February 2015, project members reformatted the source code of the versions supported at the time (0.9.8 through 1.0.2) to ensure style consistency throughout the project.

% \begin{figure}[ht]
%     \centering
%     \includegraphics[width=\linewidth]{code-style.pdf}
%     \caption{Coding Style Models}
%     \label{fig:code_style}
%     \Description{Plots of RDD models of coding style metrics}
% \end{figure}

\begin{table*}

\caption{\label{tab:code_style}Code Style and Language Feature Regression Discontinuity Models}
\centering
\begin{tabular}[t]{lrrrrrrrr}
\toprule
\multicolumn{1}{c}{} & \multicolumn{2}{c}{Style Inconsistency (1.00)} & \multicolumn{2}{c}{Goto Density (0.86)} & \multicolumn{2}{c}{Preprocessor (0.98)} & \multicolumn{2}{c}{Conditional (0.98)} \\
\cmidrule(l{3pt}r{3pt}){2-3} \cmidrule(l{3pt}r{3pt}){4-5} \cmidrule(l{3pt}r{3pt}){6-7} \cmidrule(l{3pt}r{3pt}){8-9}
  & Estimate & Std Error & Estimate & Std Error & Estimate & Std Error & Estimate & Std Error\\
\midrule
Pre-Intervention Level & 57.045\rlap{\textsuperscript{***}} & 0.276 & 0.07\rlap{\textsuperscript{***}} & 0 & 0.168\rlap{\textsuperscript{***}} & 0.001 & 0.014\rlap{\textsuperscript{***}} & 0\\
Change in Level & -52.957\rlap{\textsuperscript{***}} & 0.377 & 0.002\rlap{\textsuperscript{***}} & 0 & -0.016\rlap{\textsuperscript{***}} & 0.001 & -0.003\rlap{\textsuperscript{***}} & 0\\
Pre-Intervention Slope & -0.09\rlap{\textsuperscript{***}} & 0.019 & 0 & 0 & 0\rlap{\textsuperscript{***}} & 0.000 & 0 & 0\\
Change in Slope & 0.103\rlap{\textsuperscript{***}} & 0.026 & 0 & 0 & 0\rlap{\textsuperscript{***}} & 0.000 & 0 & 0\\
\bottomrule
\multicolumn{9}{l}{RDD models show changes in style and language feature metric means and slopes at February 2015. Each model's $R^2$ metric}\\
\multicolumn{9}{l}{is shown in parentheses after metric name. Statistical significance is indicated by *** p < 0.001, ** p < 0.01, * p < 0.05}\\
\end{tabular}
\end{table*}

The {\it cqmetrics} tool computes a coding style inconsistency metric based on $n = 19$ style rules selected from sources such as the Google, FreeBSD, and GNU coding style documents~\cite{spinellis2016evolution}. For each way to format a particular C construct, such as placing a space after the {\tt while} keyword, {\it cqmetrics} computes the sum of how many times the formatting rule is followed, $a_i$, and the sum of how many times the rule is not followed, $b_i$. The style inconsistency is the ratio of the smaller of two sums with the total times the rule could be applied in either way.
\begin{displaymath}
    SI = \frac{ \sum^{n}_{i=1} min(a_i, b_i) }{ \sum^{n}_{i=1} a_i + b_i } 
\end{displaymath}

This style inconsistency metric shows a dramatic drop immediately after the reformatting, from 57.045 to 4.088, as can be seen in Figure~\ref{fig:code_style}. The slope prior to reformatting was slightly downwards, starting in mid-2014 after Heartbleed, while the slope after reformatting is slowly upwards, showing a gradual increase in deviations from the coding style as new contributions are made. 
%As readers may be unfamiliar with the style inconsistency metric, we compare style consistency of OpenSSL with GnuTLS, which has a mean style inconsistency of 7.8 after Heartbleed.

As part of coding style, we examine the use of both {\tt goto} and the C preprocessor. The density of {\tt goto} statements does not change substantially after the code reformatting and cleanup in 2015. However,
the density of preprocessor statements decreases by about 10\% from 0.168 to 0.152 at the same point, with potentially problematic conditional preprocessor statements decreasing 23\% from 0.014 to 0.011. 

The substantial improvements in coding style consistency and reduction in C preprocessor use described above suggest that the OpenSSL project focused on improving the readability and maintainability of their code after Heartbleed.

\subsection{Project Activity}

\rq 5 addresses project activity. The OpenSSL project has become much more active since the discovery of Heartbleed. Contributors made 13,238 git commits (52.7\%) in the 68 months after Heartbleed. Only 11,905 commits (47.3\%) were made in the 197 months before Heartbleed. The difference in the number of commit authors is even larger, with only 51 commit authors (8.8\%) before Heartbleed and 554 authors (96.0\%) after Heartbleed. Note that the percentages do not sum to 100\%, as a small subset of authors were active both before and after Heartbleed. 

The tenfold growth in the number of unique authors does not equate to a tenfold growth in the number of commits after Heartbleed. This can be explained in part by the substantial number of authors who contributed only a single commit. Only ten (19.6\%) of the 51 pre-Heartbleed authors contributed just one commit. After Heartbleed, 331 (59.7\%) of the 554 authors contributed only a single commit. The large growth in both total and single commit authors suggests that the OpenSSL project has become much more attractive to outside contributors after Heartbleed.

In addition to looking at total numbers of commits and authors, we perform a time series analysis of the number of commits and the number of authors per month. We compute the number of authors per month by counting the number of unique authors who appear in {\tt git} commit author fields during a month. Authors are identified by name, not by e-mail address. We fit RDD models to both time series.

Project activity begins changing immediately after Heartbleed, so we use the month of Heartbleed (April 2014) as the cutoff date. Due to the high variance in both of these metrics after Heartbleed, we use a bandwidth of 50 months on each side of the Heartbleed month rather than 25 months used above. This high variance is reflected in the lower adjusted $R^2$ values for these models compared to most of our previous models, as shown in Table~\ref{tab:activity}. Figure~\ref{fig:project_activity} shows the model fit and data points. While project activity occurs in spurts, there are clear and substantial changes in activity after Heartbleed. The change in number of commits per month (122.44) is much larger than the number of monthly commits before Heartbleed (46.22). Similarly, the change in number of unique authors per month (13.65) is much larger than the number of monthly authors before Heartbleed (7.06).

% \begin{figure}[ht]
%     \centering
%     \includegraphics[width=\linewidth]{project-activity.pdf}
%     \caption{Project Activity Models}
%     \label{fig:project_activity}
%     \Description{Plots of RDD models of project activity}
% \end{figure}

\begin{table*}

\caption{\label{tab:activity}Project Activity Regression Discontinuity Models}
\centering
\begin{tabular}[t]{lrrrr}
\toprule
\multicolumn{1}{c}{} & \multicolumn{2}{c}{Commits per Month (0.51)} & \multicolumn{2}{c}{Unique Authors per Month (0.80)} \\
\cmidrule(l{3pt}r{3pt}){2-3} \cmidrule(l{3pt}r{3pt}){4-5}
  & Estimate & Std Error & Estimate & Std Error\\
\midrule
Pre-Intervention Level & 46.22\rlap{\textsuperscript{*}} & 21.801 & 7.057\rlap{\textsuperscript{***}} & 1.416\\
Change in Level & 122.438\rlap{\textsuperscript{***}} & 31.461 & 13.653\rlap{\textsuperscript{***}} & 2.043\\
Pre-Intervention Slope & -0.23 & 0.751 & 0.108\rlap{\textsuperscript{*}} & 0.049\\
Change in Slope & 1.904 & 1.079 & 0.039 & 0.070\\
\bottomrule
\multicolumn{5}{l}{Models show changes in metric means and slopes at April 2014. Each model's $R^2$ is shown in}\\
\multicolumn{5}{l}{parentheses after metric name. Significance indicated by *** p < 0.001, ** p < 0.01, * p < 0.05}\\
\end{tabular}
\end{table*}

\subsection{Software Engineering Practices}

\rq 6 focuses on OpenSSL's software engineering practices. In June 2014, shortly after Heartbleed, the OpenSSL project published a project roadmap~\cite{openssl_roadmap_2014}, identifying current issues, objectives, and forthcoming features. Project issues included a backlog of bug reports, some of which had been open for years, incomplete and incorrect documentation, code complexity, inconsistent coding style, a lack of code review, and the absence of a release strategy and a security policy. The developer we interviewed indicated that the OpenSSL team did not use code metrics to direct their software engineering efforts.

The CII began funding OpenSSL in 2014, enabling the developer team to grow rapidly. By December 2014, team size had increased from two main developers to fifteen project members and four full time funded developers~\cite{rsa16}. The project began formalizing decision making and published a vulnerability handling policy. Project development moved to GitHub. According to the developer we interviewed, the motivation for using GitHub was to increase transparency and attract more developers.

OpenSSL published its first release strategy in December 2014, establishing end of life dates and planning for future versions. Version 1.0.2 was created as a long term support release with a backwards compatible API, while version 1.1.0 was planned to improve design and code while breaking compatibility with earlier versions.
% OpenSSL published its first release strategy in December 2014. The release strategy established end of life dates for existing releases of OpenSSL, changed the version numbering system so letter versions only contained bug fixes, and described plans for future versions. Version 1.0.2 was created as a long term support release with a backwards compatible API, while version 1.1.0 was planned to improved design and code while breaking compatibility with earlier versions. The CII decided to perform its code audit on 1.1.0.

The project published bylaws in February 2017, describing project roles, including the OpenSSL Management Committee (OMC) and Committers~\cite{openssl_bylaws}. Starting in July 2014, the OpenSSL project required that code submissions be reviewed and approved by a core team member. The updated 2017 committer policy required that all code submissions to be reviewed and approved by at least two committers, one of whom must be an OMC member~\cite{openssl_committers}.
%The bylaws were accompanied by another document, describing guidelines for committers, including new code review requirements. 

The CII's Best Practices Badge project provides a guide to good open source development practices. The badge project was established soon after Heartbleed to provide a method for open source developers to certify that their projects follow best practices. Attainment of a badge requires meeting 66 criteria in six categories: basics, change control, reporting, quality, security, and analysis~\cite{ciibadge_criteria}. 

Prior to the changes made in response to Heartbleed, OpenSSL had completed 62\% of badge requirements. The OpenSSL project attained its CII badge in February 2016, by enabling TLS for its web site, protecting downloads of OpenSSL with TLS, publishing processes for reporting vulnerabilities and contributing code, using static and dynamic analysis before public releases, and using continuous integration.

%Requirements range from having a TLS encrypted method for downloading software to using static and dynamic analysis tools.

% Projects are rated on the criteria using a tiered percentage scale, ranging up to 300\%. A badge can be attained with a tiered percentage score of 100\%, while scores of 200\% and 300\% are needed for the silver and gold badges. Out of over 2800 projects that began the badge application process by the end of 2019, there are 390 badged projects, fourteen of which have silver badges and three of which have gold badges.

% Prior to the changes made in response to Heartbleed, OpenSSL had a badge percentage score of 62\%. The OpenSSL project attained its CII badge in February 2016, with a score of 105\%. OpenSSL improved its badge score during this two year period by enabling TLS for its web site, protecting downloads of OpenSSL with TLS, publishing processes for reporting vulnerabilities and contributing code, using static and dynamic analysis before public releases, and using continuous integration.
% https://bestpractices.coreinfrastructure.org/en/projects/87 (pre-HB)
% https://bestpractices.coreinfrastructure.org/en/projects/54 (post-HB)
% Added: https website, provides contribution requirements, vulnerability report process, unit test policy for new functionality, https downloads, static and dynamic analysis, CI

% git log --reverse .travis.yml
The OpenSSL project began using Travis for continuous integration in August 2015. In additional to building the software and performing unit tests, OpenSSL's Travis configuration reports test coverage to {\tt coveralls.io} and runs the {\tt flake8} static analysis tool on the library's python scripts. Continuous integration has been shown to change other software development practices, such as code contribution processes, issue handling, and testing~\cite{zhao2017impact}. After adopting continuous integration, OpenSSL changed in all of those areas, adopting a new code contribution policy, migrating issue tracking from Request Tracker to GitHub~\cite{openssl_issues}, and increasing unit testing.

\begin{figure}[!ht]
    \centering
    \includegraphics[width=\linewidth]{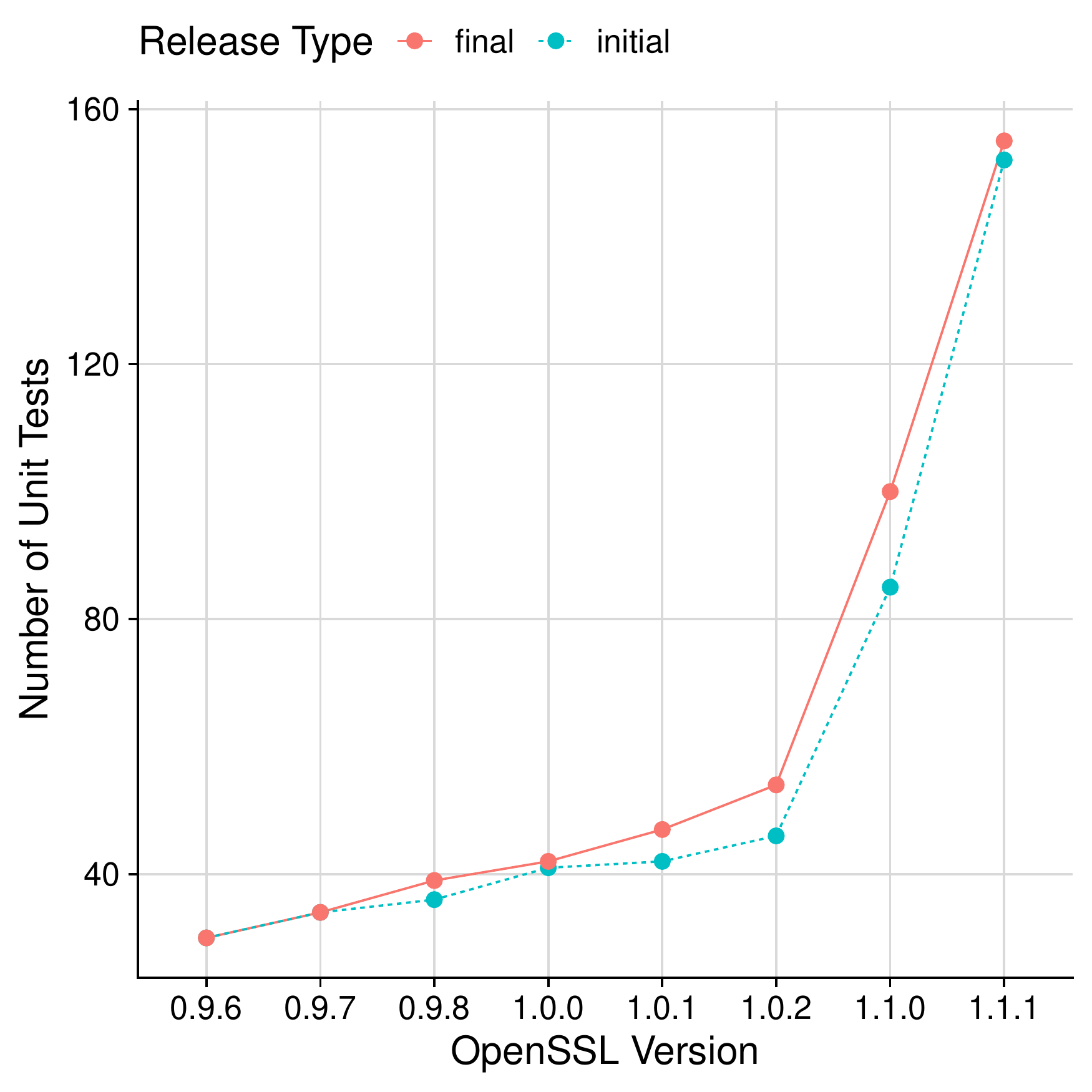}
    \caption{OpenSSL Unit Tests by Version}
    \label{fig:unit_tests}
    \Description{Plot of OpenSSL unit tests by version}
\end{figure}

OpenSSL testing practices changed considerably after Heartbleed. The project adopted a new framework to make it easier to write unit tests. The number of unit tests increased from 42 in version 1.0.1 (the most recent version as of Heartbleed) to 152 in 1.1.1 (the current version). Figure~\ref{fig:unit_tests} displays the growth in unit tests by major version, including the trend for the number of tests at the time of initial release and the number of tests for the most recent release. It is important to note that the time between initial and final release of a version is much larger for older versions. Version 0.9.8 adds three tests from first to final release, which happened over the course of ten years and 35 minor releases. Version 1.1.1 also adds three tests, but over a much shorter time period: three minor releases and a single year. Modern versions (1.1.0 and 1.1.1) added approximately one test per minor release, while older versions rarely added tests for minor releases.

The OpenSSL project began measuring test coverage in 2016. Test coverage grew from 54.6\% in 2016 to 64.2\% in 2019 as measured by {\tt coveralls.io}. There is no coverage data prior to 2016, but the increase in the number of unit tests from the last pre-Heartbleed version (1.0.1), which had 47 unit tests, to the version released in 2016 (1.1.0), which had 85 unit tests on initial release, suggests that code coverage was lower before test coverage statistics were collected.

OpenSSL has incorporated fuzz testing in its unit tests, and Google's OSS Fuzz project~\footnote{\url{https://google.github.io/oss-fuzz/}} also tests OpenSSL. While only sixteen OpenSSL vulnerabilities identify the technique or tool used to find the vulnerability, all sixteen vulnerabilities identify fuzz testing as the technique. Nine of the sixteen identify OSS Fuzz as the tool used, while five identify libFuzzer and two identify TLS-Attacker.

The OpenSSL project's code is regularly scanned by a security-oriented static analysis tool as part of the Coverity Scan project~\footnote{\url{https://scan.coverity.com/}}. When OpenSSL was initially scanned in 2006, commit messages reported fixing several bugs reported by Coverity. Coverity related commit messages continue through January 2, 2009. The next mention of Coverity in a commit message occurs on May 5, 2014, a month after Heartbleed. Similar commit messages continue through the end of 2019, so static analysis has continued to be used since 2014. Our interviewee indicated that Coverity scan results are primarily used when preparing a new release of OpenSSL.

% the locus of the first development activity by immigrants is strongly determined by modularity, complexity,etc.of the target file or class. Community, joining,and specialization in open source software innovation: a case study.

\subsection{Sustainability of Changes}

Our final research question, \rq 7, focuses on the durability of the changes made in response to Heartbleed. Improvements in the OpenSSL project during the immediate post-Heartbleed era can be seen in the project's attainment of a CII Best Practices badge. The badge score improved from 62\% to 105\%. While the project has not updated its badge application since 2016, we were able to use the application to manually verify that OpenSSL has continued to follow the best practices identified in its application, including continuous integration, static analysis, and fuzz testing.

When we examine the evolution of code metrics, we find that most improvements have been sustained. Code complexity, as measured by nesting depth, cyclomatic complexity, and Halstead complexity, has continued to decline since 2016. The maximum nesting depth remains at the level it dropped to after the code cleanup. Style inconsistency has begun to grow at a slow rate, but the December 2019 value of 7.09 is still far below the April 2014 value of 58.9. 

\begin{figure}[ht!]
    \centering
    \includegraphics[width=0.8\linewidth]{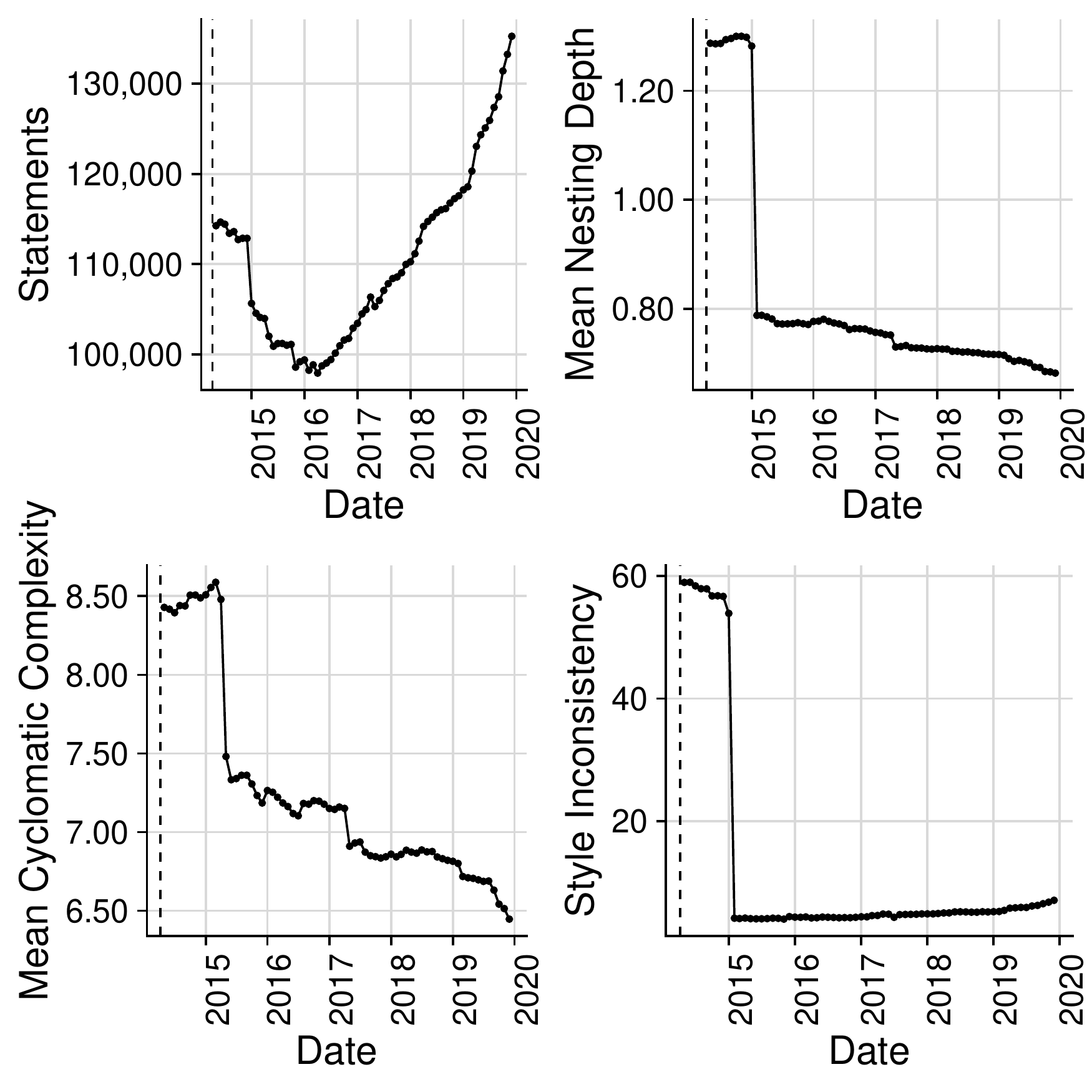}
    \caption{Four Code Metrics after Heartbleed}
    \label{fig:sustainability}
    \Description{Time series graph of code metrics 2014-2019}
\end{figure}

Growth of the code base resumed in 2016, as features like TLS 1.3 were added, showing that code shrinkage inspired by Heartbleed was a temporary change in software evolution. Figure~\ref{fig:sustainability} shows the post-Heartbleed evolution of a sample of metrics, including code size in statements, mean nesting complexity, mean cyclomatic complexity, and style inconsistency.

The security audit commissioned by the CII was performed on version 1.1.0 of OpenSSL in 2016 by the Open Crypto Audit Project (OCAP)~\cite{openssl11_audit}. The audit included both code review and fuzz testing. The audit reported several potential bugs in OpenSSL, including two possible code execution vulnerabilities and two possible denial of service (DoS) vulnerabilities. Other issues were reported by the audit team as low severity or difficult to exploit. Only one OpenSSL vulnerability report identifies the OCAP audit team as the reporter.
%Only one vulnerability for OpenSSL (CVE-2016-2181) identifies the OCAP audit team as the reporter.

The Open Source Technology Improvement Fund (OSTIF) performed an audit of OpenSSL 1.1.1, with a focus on the new TLS 1.3 protocol and changes made to the Pseudo Random Number Generator (PRNG)~\cite{openssl111_audit}. Their report was published in 2019. The OSTIF audit combined manual code review with fuzz testing. OSTIF found two DoS vulnerabilities prior to the release date, enabling the OpenSSL team to fix them before release. The report identified some areas where code quality could be improved by checking return values and implementing global checks for NULL values.

\section{Threats to Validity}

\textit{Construct Validity:} We use multiple widely used metrics for measuring code size and complexity to avoid bias in measuring those code characteristics. The style inconsistency metric we use has been used in other studies too~\cite{spinellis2016evolution}. 

\textit{Internal Validity:} One of the most important validity threats to RDD time series analysis is the presence of an event near the cutoff date that influences the observed changes. We mitigated this threat by thoroughly examining OpenSSL blog entries, commit messages, and email archives in the months before and after our cutoff dates for such events. Another threat to internal validity is that only one OpenSSL developer consented to be interviewed, which may bias our qualitative data.
% See Hausman & Rapson, Regression Discontinuity in Time for more details
% if we write a longer version of this paper.
%other than the Heartbleed vulnerability used for the project activity cutoff date and the code cleanup used for the code metric cutoff date.

\textit{External Validity:} As this work is a case study of a single project, we cannot generalize our conclusions to other open source projects. We therefore leave the question of how open source projects can successfully react to security events to future work.

% See A Practical Guide to RDD, page 6, threats to internal validity

\section{Conclusions}

The Heartbleed vulnerability brought dramatic changes to OpenSSL, transforming an almost moribund project to an active project with substantial improvements in code quality and security. OpenSSL remains the most commonly used TLS library on public web servers five years after Heartbleed, according to IPv4 scan data collected by \url{censys.io}. These improvements provide a model for how open source projects can adapt and improve after a major security event.

We found substantial and sustained improvements in code quality. Code complexity declined sharply during the major code cleanup activity in 2015, and both cyclomatic and Halstead complexity have continued to decline. The code cleanup made coding style much more consistent. While style inconsistency has slowly increased since the cleanup, it remains much lower than before Heartbleed.

The number of vulnerability reports dramatically increased for three years after Heartbleed before returning to previous levels. Only five of the 91 vulnerabilities found in those three years were in post-Heartbleed code, so this represents a substantial improvement in security.
Positive results from two external code audits also suggest that the security of OpenSSL has greatly improved. This means that vulnerability count is {\bf not} a useful indicator of project security. Low vulnerability counts may just indicate that a project is devoting little effort to finding vulnerabilities.

Our results suggest that project activity and practices {\it may} be better predictors of project security. The CII badge {\it may} be a good indicator of project security, since it requires good open source development practices. To understand how generalizable these recommendations are, we plan to compare the software evolution of OpenSSL with that of related projects, such as BoringSSL and GnuTLS. We also plan to examine the impact of the CII badge and project activity on the development practices and security across multiple open source projects to validate these ideas.

We have provided a replication package for this paper~\cite{msr2020replication}, which includes the software metrics and project activity data used in this paper. The package includes data collection scripts and the code used to create the models and generate the figures and tables for this paper. Documentation on how to use the scripts is also provided.

% Bibliography
\bibliographystyle{ACM-Reference-Format}
\bibliography{openssl-case-study}
%
% The use of \BibTeX\ for the preparation and formatting of one's
% references is strongly recommended. Authors' names should be complete
% --- use full first names (``Donald E. Knuth'') not initials
% (``D. E. Knuth'') --- and the salient identifying features of a
% reference should be included: title, year, volume, number, pages,
% article DOI, etc.

\end{document}